\documentclass[prl,twocolumn,notitlepage,amsfonts,amssymb,amsmath,floatfix,tightenlines,superscriptaddress]{revtex4-2}

\usepackage{amsfonts,amssymb}
\usepackage{amsmath,times}
\usepackage{graphicx}
\usepackage[colorlinks=true,linkcolor=blue,anchorcolor=red,citecolor=blue,urlcolor=blue]{hyperref}
\usepackage{color}
\usepackage{cleveref}
\usepackage[lofdepth,lotdepth,caption=false]{subfig}

\newcommand{\vect}[1] {\mathbf{#1}}

\newcommand{\xik}{\xi_{\mathbf{k}}}

\newcommand{\vecq}{\vect{q}}

\newcommand{\dif} {\mathrm{d}}  

\newcommand{\FT} {{\Sigma_2}}
\newcommand{\GT} {{\Sigma_1}}

\newcommand{\Tmat} {t}
\newcommand{\AT} {A}

\newcommand{\mupair}{ \mu_{\text{pair}} }

\begin{document}

\title{Universal approach to light driven ``superconductivity" via preformed pairs}

\author{Ke Wang}
\email{kewang07@uchicago.edu}
\affiliation{Department of Physics and James Franck Institute, University of Chicago, Chicago, Illinois 60637, USA}
\author{Zhiqiang Wang}
\affiliation{Hefei National Research Center for Physical Sciences at the Microscale and School of Physical Sciences, University of Science and Technology of China,  Hefei, Anhui 230026, China}
\affiliation{Shanghai Research Center for Quantum Science and CAS Center for Excellence in Quantum Information and Quantum Physics, University of  Science and Technology of China, Shanghai 201315, China}
\affiliation{Hefei National Laboratory, University of  Science and Technology of China, Hefei 230088, China}
\affiliation{Department of Physics and James Franck Institute, University of Chicago, Chicago, Illinois 60637, USA}
\author{Qijin Chen}
\affiliation{Hefei National Research Center for Physical Sciences at the Microscale and School of Physical Sciences, University of Science and Technology of China,  Hefei, Anhui 230026, China}
\affiliation{Shanghai Research Center for Quantum Science and CAS Center for Excellence in Quantum Information and Quantum Physics, University of  Science and Technology of China, Shanghai 201315, China}
\affiliation{Hefei National Laboratory, University of  Science and Technology of China, Hefei 230088, China}
\author{K. Levin}
\affiliation{Department of Physics and James Franck Institute, University of Chicago, Chicago, Illinois 60637, USA}
\date{\today}

\begin{abstract}
  While there are many different mechanisms which have been proposed
  to understand the physics behind light induced ``superconductivity",
  what seems to be common to the class of materials in which this is
  observed are strong pairing correlations, which are present in the
  normal state.  Here we argue, that the original ideas of Eliashberg
  are applicable to such a pseudogap phase and that with exposure to
  radiation the fermions are redistributed to higher energies where
  they are less deleterious to pairing.  What results then is a
  photo-induced state
  with dramatically enhanced number of nearly condensed fermion pairs.
  In this phase, because the a.c. conductivity,
  $\sigma(\omega) = \sigma_1(\omega) + i \sigma_2(\omega)$, is
  dominated by the bosonic contribution, it can be computed using
  conventional (Aslamazov Larkin) fluctuation theory.  We, thereby,
  observe the expected fingerprint of this photoinduced
  ``superconducting" state which is a $1/\omega$ dependence in
  $\sigma_2$ with fits to the data of the same quality as found for
  the so-called photo-enhanced (Drude) conductivity scenario. Here,
  however, we have a microscopic understanding of the
  characteristic low energy scale which appears in transport and
  which is necessarily temperature dependent. This approach also
  provides insight into  recent observations of concomitant
  diamagnetic fluctuations.
  Our calculations suggest that the observed light-induced phase in these
  strongly paired superconductors has only short range phase coherence
  without long range superconducting order.
\end{abstract}

\maketitle

\section{Introduction}
The phenomenon of light induced phase and
structural transitions~\cite{Mankowsky2014,Tang2023}
and more specifically light induced, transient superconductivity well
above $T_c$, has generated enormous excitement in the
community~\cite{Orenstein2012,Nicoletti2016,Cavalleri2018,Demsar2020,De2021}.  In these
latter experiments, a very fast, laser pulse of sub-picosecond (ps) duration
is applied, and with a delay of around 1 to 10 ps the system is probed and
the conductivity at mid-infrared and terahertz frequencies is
measured.  What is observed are superconducting-like signatures
appearing in the normal state.  Most notable among these is a low
frequency upturn in the imaginary conductivity $\sigma_2(\omega)$, not
so different from the $1/\omega$ dependence of a true superconductor.

In this paper we argue, as has Uemura~\cite{Uemura2019}, that, while
there are many different mechanisms
~\cite{Nava2018,Dolgirev2022,Salvador2024,Knap2016,Babadi2017,
Raines2015,Kim2016,Komnik2016,Okamoto2016,Sentef2017,Coulthard2017,Kennes2017,Mazza2017,Murakami2017,Sentef2015,Dasari2018,Wang2018,Tindall2020}
suggested to explain these transient superconducting-like properties
of matter, there seem to be important experimental commonalities.  In
particular, these transient light induced indications of
superconductivity are observed at often high temperatures associated
with a normal state that contains preformed pairs
~\cite{Okamoto2016,Jotzu2023,Cavalleri2018}.

The goal of this paper is to build on these commonalities and to
propose a more universal mechanism for such light induced phenomena,
with an emphasis on addressing transport signatures in more
quantitative detail.  Here we note that in a strong pairing scenario
such as we consider~\cite{Chen2024,Chen2005}, we should view these
preformed pairs as necessarily co-existing with (pseudo)gapped
fermions.  The fermions which are present along with preformed pairs
play a central role in understanding these experiments, as they
provide an important handle for implementing the physics of a highly
non-equilibrium Eliashberg-like scenario~\cite{Eliashberg1970} for
light-enhanced superconductivity.

\begin{figure*}
\centering
\includegraphics[width=6.2in,clip]
{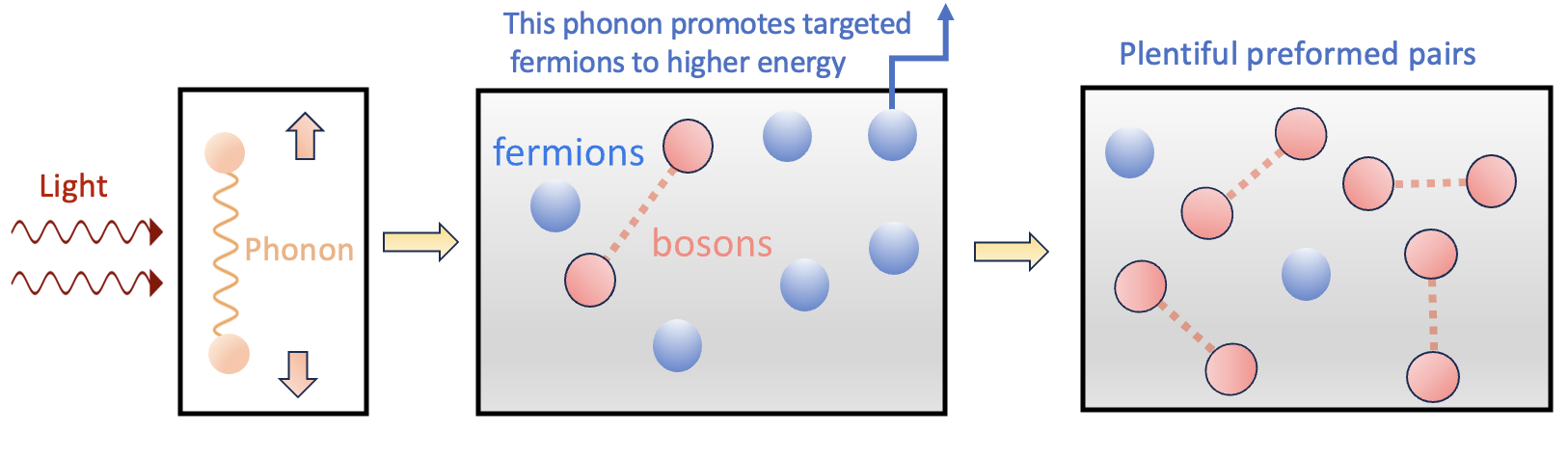}
\caption{This figure represents a simple physical picture of the Eliashberg mechanism
now extended to the
normal phase. The effects of light (indirectly through a specially selected
phonon) 
serve to redistribute those particular fermions associated with superconductivity to 
higher energies
\cite{Tredwell1975,Tredwell1976},
As a consequence, after radiation,
while the fermion number is
substantially reduced,
the pairing is strengthened and preformed pairs (bosons) become more plentiful,
thus leading to enhanced fluctuation transport signatures.}
\label{fig:1}
\end{figure*}

The Eliashberg picture begins with the premise that radiation promotes
the gapped fermionic quasiparticles to higher energies, which
is associated with
a larger pairing gap.
We have no well established
tools for treating such highly non-equilibrium behavior, but it is
claimed that there is no doubt that the Eliashberg effect
exists~\cite{Klapwijk2020}.  What we address here is its application
and generalization to the normal (non-superconducting) phase. In the
Eliashberg scenario below $T_c$ one views the quasiparticles as out of
equilibrium with condensed Cooper pairs.  Similarly, for the above
$T_c$ counterpart one views the quasiparticles as out of equilibrium
with preformed pairs.  It was previously found that in the Eliashberg
scenario there is a substantial
enhancement~\cite{Robertson2009,Heslinga1993} of the transition
temperature $T_c$.  For the normal state in our theory we find a
substantial enhancement of the counterpart pairing onset temperature
$T^*$.  A consequence is that, as the number of relevant fermionic
excitations effectively decreases, this leads to an increase in both
the gap size and the boson number, which will, in turn, have important
implications for the normal state transport properties.

An important component of these experiments \cite{Cavalleri2018} are 
resonant phonons which initiate these ``superconducting"-like
responses to light. Exciting these particular phonons is important
presumably because they serve to select the specific \textit{fermions}
which participate in the pairing. We argue that what is responsible
for facilitating this non-equilibrium fermion redistribution is not
the direct interaction between the photons and fermions, which has
less specificity, but rather the direct interaction between the
fermions and selected phonons. Indeed, an extended version of
Eliashberg theory in which the enhancement mechanism is based on
phonon absorption is now well
established~\cite{Tredwell1975,Tredwell1976}. Adding support is the
observation that, for example in the cuprates, the transition
temperature seems to depend rather sensitively on the oxygen apical
mode ~\cite{Johnston2010}. Similarly, exciting this mode has been
found to be effective in initiating the light-enhanced behavior.

The physical picture associated with the Eliashberg
mechanism in the normal state is schematically represented in
Fig.~\ref{fig:1}.
As has been pointed out~\cite{Robertson2009}
one can interpret this mechanism more
microscopically as being associated with a radiation-induced increase in the
strength of the attractive interaction.
It is important to emphasize that this increase 
in the attraction only occurs by redistribution of fermions.
It is not related to the microscopic details of the pairing mechanism which leads to the
superconductivity. Because it occurs in the normal state, this leads to
enhanced signatures of precursor (fluctuation) superconductivity.

Our recent Review~\cite{Chen2024} provided an overview of
these strongly paired superconductors.
Interestingly among materials which exhibit light enhanced ``superconductivity"
were 
\begin{enumerate}
\item Members of the cuprate family
~\cite{Cavalleri2018,Nicoletti2016}.  
\item  Members of the Fe-Se family~\cite{Isoyama2021}.
\item A particular organic superconductor~\cite{Buzzi2021}, known as
$\kappa$-BEDT-TTF.
\item  A particular fulleride 
~\cite{Cavalleri2018,Mitrano2016} K$_3$C$_{60}$.
\end{enumerate}
All of these we identified as strongly paired except for
the fulleride, but it now appears
that this one notable example (K$_3$C$_{60}$) 
also seems to belong to this category~\cite{Ren2020}.
We found
additional
strong pairing superconductors belonging to the class of artificial materials~\cite{Nakagawa2021,Park2021}
which have not been addressed in pump-probe experiments. Otherwise the overlap
between those we identified and
those which exhibit this transient
superconductivity is surprisingly good.

\vskip3mm
\section{Preformed Pairs in Generalized BCS Theory}
The strongly-paired superconductor scenario~\cite{Chen2005,Chen2024}
depends on an enhanced pairing strength, beyond that associated with
the BCS regime.  There are two characteristic temperatures in such a
superconductor: $T^*$ and $T_c$.  Here $T^*$, introduced earlier,
represents the onset temperature for the opening of a gap, while $T_c$
represents the ordering or phase coherence temperature. The
excitations are pseudogapped fermions (having a Bogoliubov-like
dispersion) as well as non-condensed or preformed pairs at $T<T^*$.
The gap parameter $\Delta$ takes on a different character above and
below $T_c$.  Above $T_c$, which is of primary interest here,
$\Delta \equiv \Delta_{\text{pg}}$, where ``pg" refers to the
pseudogap.

To understand the nature of preformed pairs within a generalized BCS formalism
it is useful to recognize
that the BCS gap equation can be written compactly in terms of the Gor'kov
function $F(k) \equiv \Delta G(k)G_0(-k)= \Delta /(\omega_n^2 + E_{\vect k}^2)$, 
where $G(k)=( i \omega_n + \xi_{\vect k} )/ ( - \omega_n^2 - E_{\vect k}^2)$ is the usual
(diagonal) Gor'kov Greens function,
and $E_{\vect k} = \sqrt{\xik^2 + \Delta^2(T)}$ is the gapped fermionic quasiparticle dispersion. 
$G_0(k)=( i\omega_n -\xi_{\vect k})^{-1}$ and $k=(i\omega_n ,\vect k)$, with $\omega_n=(2n+1)\pi T$ the fermionic Matsubara frequency. 
We have adopted the units $\hbar = k_\text{B} = c = 1$. Note that for simplicity, we drop here the order parameter symmetry 
multiplier, $\varphi_\mathbf{k}$, (which would take take the form $\varphi_\mathbf{k}  = \cos k_x - \cos k_y$ for a
systems with d-wave pairing).
Using these Greens functions we can write the gap equation as
\begin{align}
 \Delta(T)  & = U  \sum_{k} F(k)  \nonumber \\
 \Rightarrow 0  & =  \left[ -  U^{-1} +  \sum_k G(k) G_0(-k)  \right] \Delta(T) ,
\end{align}
where $U>0$ is the attraction strength and $\sum_k \equiv T\sum_n \sum_{\vect k}$. 
This form motivates a natural choice for the pair propagator
or $t$-matrix, characterizing the non-condensed  or preformed pairs,
which has been discussed in the literature~\cite{Kadanoff1961}\footnote{It should be emphasized that this asymmetric combination of $G_0G$ is derived rigorously from the equations of motion for the Green's function series and is not an ad hoc choice.}
%
\begin{equation}
\label{eq:tmatrix}
t^{-1}(q)=\sum_{k}G(k)G_{0}(q-k) -  U^{-1},
\end{equation}
where $q=(i\Omega_m,\vect{q})$ with $\Omega_m=2m \pi T$ the bosonic
Matsubara frequency, and $m\in \mathcal{Z}$.  We can view the gap
equation below $T_c$ as equivalent to the requirement that
$t^{-1} (0) = 0$ while above $T_c$, $t^{-1} (0) = Z \mupair \ne 0$
where $Z$ is a coefficient of proportionality that arises when expanding
$t^{-1}(\Omega, \mathbf{q}=0)-t^{-1}(0)$ at small $|\Omega|$. This
introduces a key parameter, $\mupair$, which represents the chemical
potential of the pairs and which identically vanishes at and below
$T_c$, as expected of a (Bose Einstein) condensation.

Thus, in the normal state we have that
\begin{eqnarray}
t^{-1}(0) =\sum_{\mathbf{k}} \frac{1-2f\left(E_{\mathbf{k}}\right)}{2E_{\mathbf{k}}}
- U^{-1}
&=&Z\mu_{{\rm pair}},
\label{eq:gap}
\end{eqnarray}
where $f(x)=1/(e^{\beta x}+1)$ is the Fermi-Dirac distribution function. 
Eq.~(\ref{eq:gap}) can be used to determine $\mupair$ provided one
knows $\Delta_{\text{ pg} }$, which can be obtained in terms of the number of
preformed pairs, $n_B$,  in the system. Here $n_B$ is to a good approximation given by
\begin{equation}
n_{B} =
\sum_{\vect{q} }b\left(\frac{\vect{q}^{2}}{2M_B}- \mupair \right)
= Z \Delta_{ \text{pg} }^2, 
\label{eq:NB}
\end{equation}
where $b(x)=1/(e^{\beta x}-1)$ is the Bose-Einstein distribution.
The pair dispersion $\vect{q}^2/2 M_B$ or equivalently the pair mass $M_B$
and all parameters in the $t$-matrix, such as $Z$,
can be obtained~\cite{Chen2024} by a small $q$ expansion of Eq.~(\ref{eq:tmatrix}). 
In determining these parameters one must similarly solve self-consistently for
the fermionic chemical potential $\mu$.
The equations to be solved are those at equilibrium which depend on the usual
Fermi distribution function
$f(E_{\mathbf{k}})$.

\begin{figure*}
\includegraphics[width=5.8in,clip]
{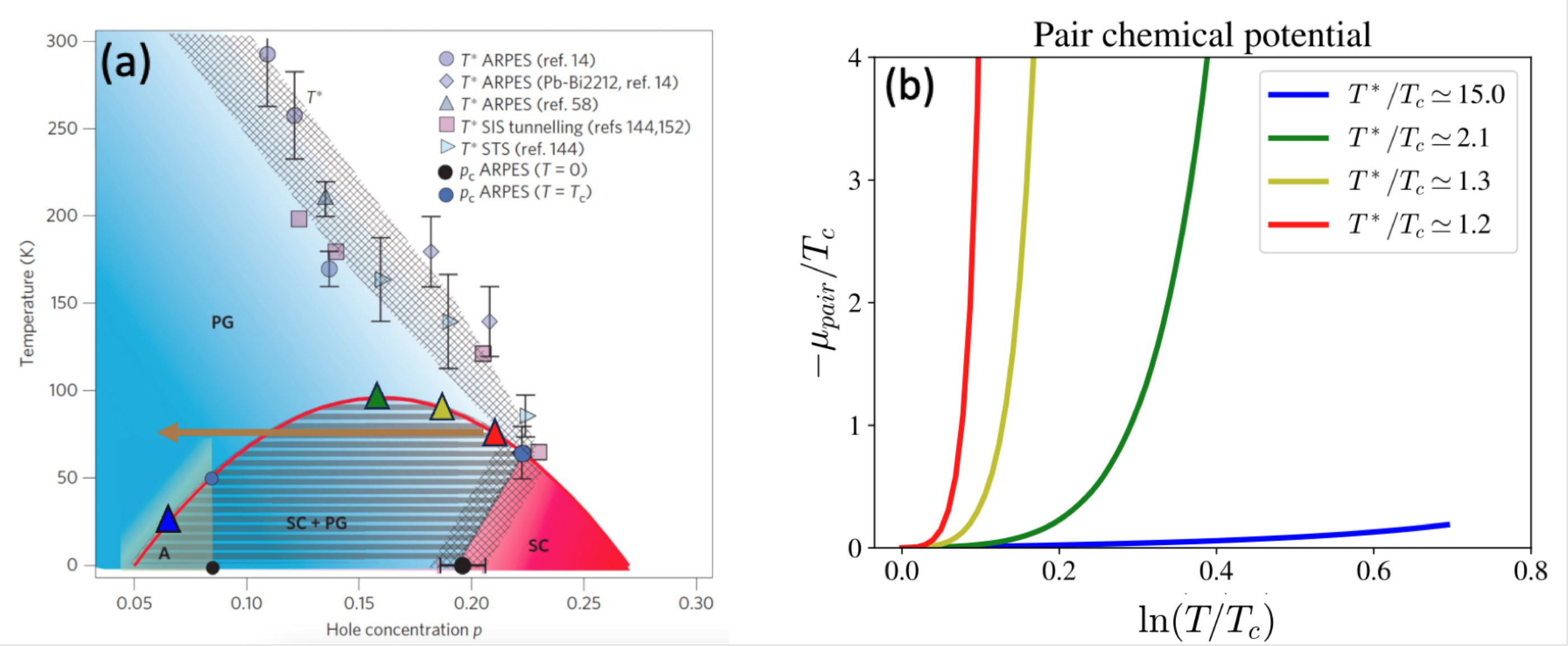}
\caption{The figure shows a prototypical  phase diagram of strongly coupled superconductors and the pair chemical potential of these states.
Panel (a) shows a prototypical phase diagram for superconductors with strong pairing, in this case for the cuprates~\cite{Hashimoto2014}.
Indicated are the $T_c$ dome and the pairing onset temperature, $T^*$.
The latter reflects the size of the attractive interaction.
This figure illustrates the effect of changing the attractive interaction (at a small $T^*/T_c$) so that, as shown by the brown arrow, a state represented by the red triangle can be transformed into a state associated with the blue triangle (at a larger $T^*/T_c$) while the temperature is held constant.
The left-hand edge of the dome corresponds to a relatively strong interaction strength.
Panel (b) shows the behavior of the bosonic chemical potential for the analogous superconductors, using the same color codes as the triangles in panel (a).
Note that systems with a larger attraction (larger $T^*/T_c$) have a substantially smaller $|\mu_{\text{pair}}|$ over a more extended temperature range.}
\label{fig:2}
\end{figure*}

\begin{figure} 
\centering
\includegraphics[width=3.3in,clip]
{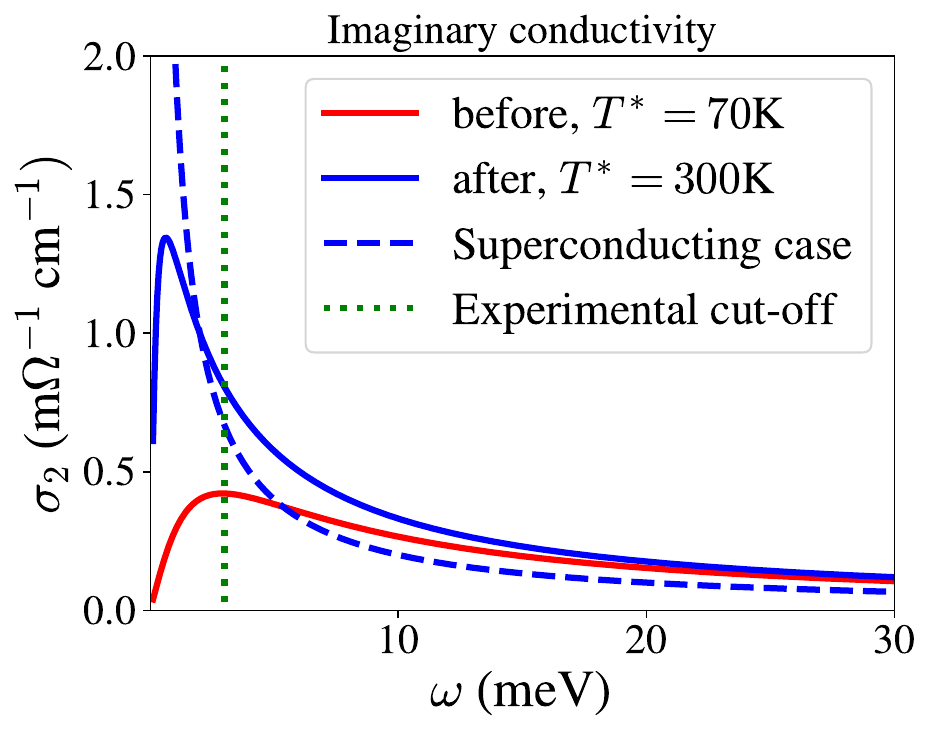}
\caption{This shows the changes in $\sigma_2(\omega)$
before (red) and after (blue) exposure to radiation. The dotted green line
indicates the lower frequency cut-off set by the experiments. 
This figure shows how the signature upturn in $\sigma_2(\omega)$ is enhanced
by the effective increase in $T^*$ associated with exposure to radiation. 
The blue line can be compared with a typical superconductor, presuming, for example
a superfluid density ratio of $n_s / n \approx 0.6 $.
}
\label{fig:3}
\end{figure}

\begin{figure}
\includegraphics[width=3.0in]
{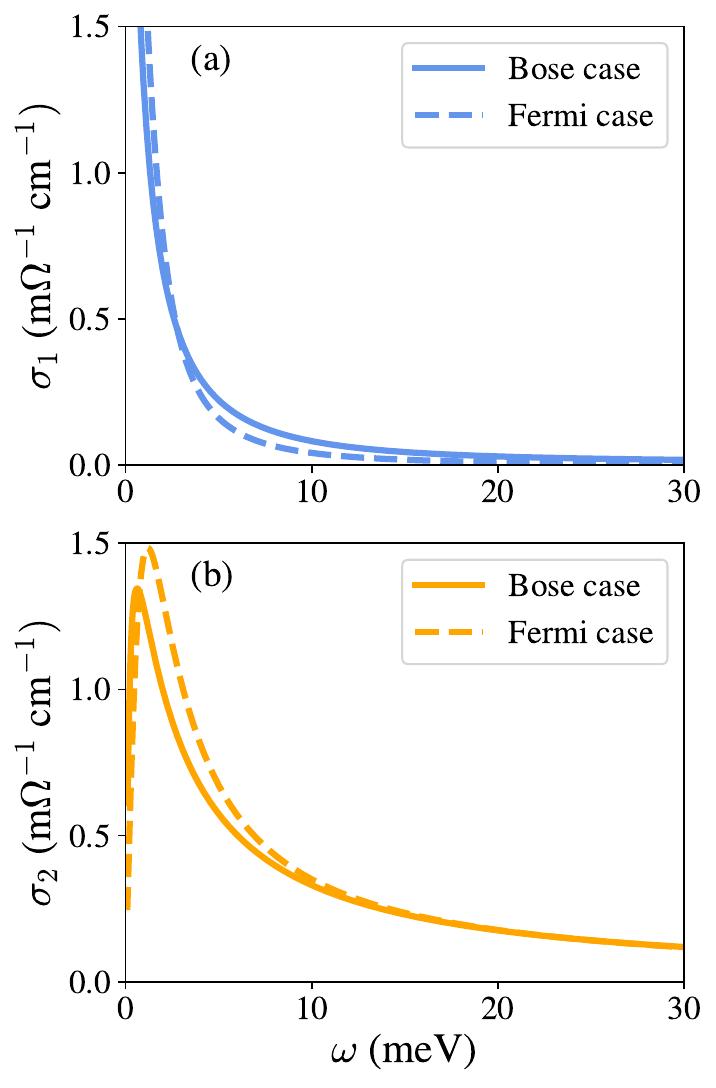}
\caption{This figure compares the Fermi (Drude) and Bose conductivities as functions of frequency.
Panel (a) shows the real part of the conductivity, and panel (b) shows the imaginary part.
When the characteristic (low) energy scales are similar, the transport properties are also comparable.
Note that in the relevant frequency range, $\sigma_1(\omega)$ is substantially smaller than $\sigma_2(\omega)$.
The assumed values in the range from 3~meV to approximately 10~meV are in reasonable agreement with experimental data.}
\label{fig:6}
\end{figure}

\begin{figure}
\includegraphics[width=3.3in,clip]
{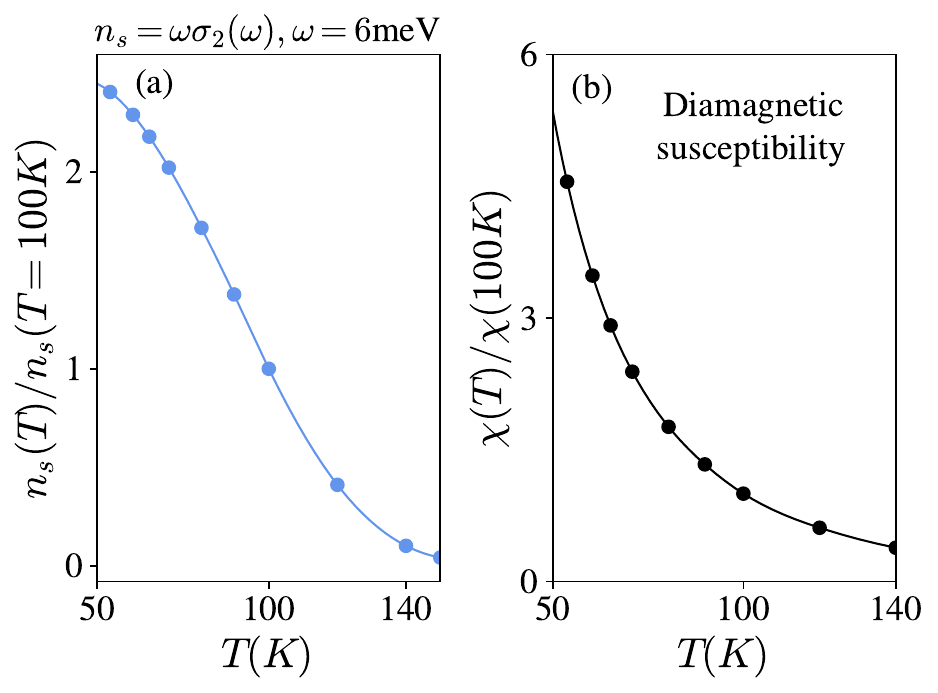}
\caption{The figure shows the temperature dependence of the effective superfluid density and the diamagnetic susceptibility.
    Panel (a) plots $\omega \sigma_2(\omega)$ at $\omega \approx 6$~meV for different initial temperatures and can be compared with Figure~5 in Ref.~\onlinecite{Liu2020}.
    Panel (b) plots the corresponding diamagnetic susceptibility for the same temperatures.
    Note that, in contrast to experimental findings, neither plot shows contributions extending up to the pairing temperature, $T^* \approx 300$~K. }
\label{fig:4}
\end{figure}


In the presence of EM radiation, we
contemplate, within the Eliashberg scenario~\cite{Eliashberg1970}, 
that there is  a redistribution of the gapped fermionic
quasiparticles to higher energies.
As a result, the key parameters $\mupair $ and $n_B$ and $\Delta_{ \text{pg} }$ as well
all are altered.
If the temperature is unchanged, such a momentum space redistribution of fermions is to be
associated with an increased $\Delta$ (and accordingly lowered chemical potential $\mu$) as the system tries to re-equilibrate. This, then, leads to stronger pairing.

To understand how, we can make some useful inferences~\cite{Robertson2009} here by
writing $$ f(E_\vect k) \rightarrow \tilde{f}(E_\vect k ) = f(E_\vect k) - \delta f (E_\vect k), $$
in the energy regime where pairing takes place. 
Here $\delta f(E_\vect k)$ represents a change in the distribution of fermions. 
We assume that, after the radiation, the system is in a quasi-equilibrated
state where the new distribution of fermionic excitations, $\tilde{f}(E_{\vect k})$, satisfies an equation similar to
Eq.~\eqref{eq:gap}, which leads to
\begin{eqnarray}
\sum_{\mathbf{k}}\frac{1-2\widetilde{f}\left(E_{\mathbf{k}}\right)}{2E_{\mathbf{k}}}
- U^{-1}
&=&Z\mu_{{\rm pair}}. \label{eq:GP2}
\end{eqnarray}
By rewriting this equation in the form of Eq.~\eqref{eq:gap} with the equilibrium $f(E_{\vect k})$, we can see that the interaction term 
$U$, which is large to begin with, is effectively strengthened by the redistribution of
the gapped fermionic excitations. These effects can be incorporated into a renormalized interaction, 
$U_{\text{eff}}$, 
given by

\begin{equation}
 \frac{1}{U_{\text{eff}}} = \frac{1}{U} - 
\sum_{\mathbf{k}}\frac{2 \delta f(E_k) }{2E_{\mathbf{k}}}
\label{eq:5}
\end{equation}
The summation in Eq.~(\ref{eq:5}) must necessarily be positive as required by the shifted
momentum space fermion redistribution. Importantly, this increase which appears in $U_{\text{eff}}$ will raise the pairing onset temperature,
$T^*$. There is no simple direct equation which focuses on this temperature, 
but it can
be approximated 
using the mean field BCS gap equation, in the limit $\Delta \rightarrow 0$.
As might be expected ~\cite{Robertson2009}, in this case the increase in $U_{\text{eff}}$
will enter exponentially as an enhancement of the characteristic temperature.
A more detailed implementation of these ideas can be found in Ref.~\onlinecite{Tikhonov2020}, albeit in the context
of a light-enhanced superconducting, rather than pseudogap phase as discussed here.
There are bounds on how large these effects can be as the number of
fermions which are redistributed can not exceed those which are initially present.

Interestingly, we can infer that the radiation exposure and subsequent
redistribution of the fermions will likely lead to
a \textit{decrease} in $T_c$, as $U_\text{eff}$ gets strongly enhanced.  This
latter, well-known ~\cite{Nozieres1985} observation is associated with
a superconducting dome-like phase diagram in which $T_c$ is
necessarily non-monotonic with varying $T^*$. The final state after radiation is to be associated with the stronger-coupling side of the $T_c$ dome in this phase diagram \cite{Chen2024}.  In a related fashion
the redistribution of fermions, now applied to the pseudogap phase,
leads as well to a \textit{decrease} in the magnitude of $|\mupair|$
and an increase in the number of bosons, $n_B$.

Importantly, for the purposes of the present paper these observations
imply that there will be an enhancement in the (bosonic) fluctuation
transport contributions. These are to be associated 
with Aslamazov-Larkin contributions and are seen to contribute
when the temperature is considerably below the gap onset temperature~\cite{Tan2004},
$T^*$. 
The signature of light-enhanced
``superconductivity" is based on the behavior of the the imaginary
contribution to $\sigma(\omega) = \sigma_1 + i \sigma_2$, where a
$1/\omega$ dependence is crucial.  We address the counterpart
calculations of $\sigma(\omega)$ in the Methods section where we
implement the known pairing fluctuation contributions to this
transport ~\cite{Larkin2005,Tan2004}.  Under radiation exposure we
expect that the number of fermion pairs is dramatically enhanced and
the bosonic contribution will dominate the ac conductivity.
This provides a normalization for $\sigma_1 (\omega)$ in
terms of the measured plasma frequency, $\omega_p$.

\section{Results}

To address this transport behavior, in the left panel of
Figure~\ref{fig:2} we present a prototypical  equilibrium phase diagram for
the behavior of $T^*$ and $T_c$, here for the case of the cuprate
superconductors \footnote{ We should point out there there are changes
  in the density across this phase diagram which are not
  relevant. Nevertheless one can see from Fig 8 in
  Ref.\onlinecite{Chen2024} that the cuprate phase diagram is rather
  typical.}.  A similar $T_c$ dome structure is seen in the organic
superconductor~\cite{Suzuki2022} which compound is also associated
with light induced ``superconductivity", but there $T^*$ was not
addressed.  In the right panel of Fig~\ref{fig:2}, we have indicated
the behavior of the pair chemical potential for a selected set of
strongly paired superconductors with the same color codes as shown in
the phase diagram.

The effects of radiation lead to a substantial increase of $T^*$. And we here presume
this happens at constant temperature. We illustrate how transport is
affected by taking as the final state
one with high $T^*$ shown by the blue triangle in the phase diagram of Fig.~\ref{fig:2}(a)
corresponding to the
blue curve in Fig.~\ref{fig:2}(b).
Any number of initial states before radiation can be contemplated, say corresponding to those
indicated by the colored triangles on the $T_c$ dome, with their respective
smaller $T^*$.
The most pronounced change in the conductivity will occur if the initial state corresponds
to the red triangle, which is the case we consider here for illustrative purposes.
This initial state before radiation, of course, is of lesser importance.


In Figure \ref{fig:3} we compare the behavior of $\sigma_2$ before and after radiation. This is to
be associated with the transition shown in Fig.~\ref{fig:2} by the brown arrow, from the red to blue curves, as $T^*/T_c$
changes from 1.2 to 15 reflecting an increase in the strength of the pairing
interaction. Also indicated in Figure \ref{fig:3} is the behavior associated with a counterpart superconductor
(blue dashed curve) and the characteristic range of frequencies accessible experimentally.
One sees that the $\sigma_2$ obtained after radiation for the preformed pairs behaves quite similarly to the superconducting $\sigma_2$
for those frequencies which are accessible to experiments.


It is important to stress that there are somewhat closely related but competing ideas in the literature
arguing that what is being observed experimentally cannot readily distinguish the superconducting
state from a non-superconductor. This is principally due to the limitation of accessible
frequencies. In this alternative scenario, 
rather than
the non-interacting bosons discussed here
one considers non-interacting fermions whose conductivity is described by Drude 
theory~\cite{Dodge2023,Dodge2023a}.
There one assumes that under exposure to radiation the fermionic lifetime
$\tau$ becomes
extremely long, thereby providing an alternative low energy scale. 
If one takes comparable
parameters for $|\mupair|$ and $\tau^{-1}$.
the frequency dependent conductivity will be very similar to the present case.

A comparison between Bose and Fermi systems, using physical units is presented in
Figure \ref{fig:6}, for both the real and imaginary components of the conductivity.
For definiteness,
in the Drude fits, previous work~\cite{Dodge2023,Dodge2023a} suggested that
$ \tau^{-1}$ might vary between 0.6 and 1.2 meV. Here we chose 1.2meV for the plots.
It is important to note that the real part vanishes more rapidly than the
imaginary term in the high frequency regime. Rather than following the power
law $1/\omega$ as in $\sigma_2$, $\sigma_1$ falls off more rapidly with $\omega$,
as the f-sum rule is exhausted by the weight of $\sigma_1$ at small frequency due to the small energy scale, $|\mupair|$ or $\tau^{-1}$. This is also compatible with light-induced experiments -- that after
radiation for a range of higher frequencies, there is very little contribution
to $ \sigma_1 (\omega)$.

Importantly, the absolute values deduced for the real and imaginary conductivities
in both theories are in reasonable agreement with the experimental values. This is
the case provided the analysis
is based on the protocol from
Ref. \onlinecite{Dodge2023} which takes into account some of the complications about
the pump-probe profile deformations which were not included in the original studies.

This raises the next question of how to distinguish between the
photoenhanced conductivity and the present preformed pair scenario.  To do so, we consider
the effects of varying temperature as well as accompanying diamagnetism
both of which play a rather important role
in the bosonic transport scenario, but do not appear as relevant for the Drude model.
Plotted on the left 
in Fig.~\ref{fig:4}
is the temperature variation of $\omega \sigma_2(\omega)$
for one particular frequency and for the same system considered in Fig.~\ref{fig:3};
here the lowest $T$ point corresponds to
$ \sigma_2(\omega) \approx \frac{ne^2}{m \omega} $, while at higher $T$
$\omega\sigma_2$ becomes progressively reduced. 
This can be compared with Figure 5 in Ref.~\onlinecite{Liu2020}.
The counterpart diamagnetic susceptibility 
$\chi_{\text{dia} } \propto - \frac{T}{|\mupair|}$,
here estimated for a quasi-two dimensional (2d) system, is plotted on the right panel for these same temperatures.

These figures show that temperature (through $\mupair$) plays a rather profound role
in bosonic fluctuation transport. This, presumably, is harder to argue for in the
alternative fermionic scenario.
Interestingly, temperature is also important in these light-driven experiments
\cite{Liu2020,Fava2024}. What is plotted in the theory here is the temperature
dependent diamagnetic susceptibility, which is somewhat different from
the experimental magnetic expulsion plot. Nevertheless, these figures make clear that the enhanced diamagnetism
found here is directly correlated with the peak structure in the optical conductivity.

\begin{figure}
\includegraphics[width=3.40in,clip]
{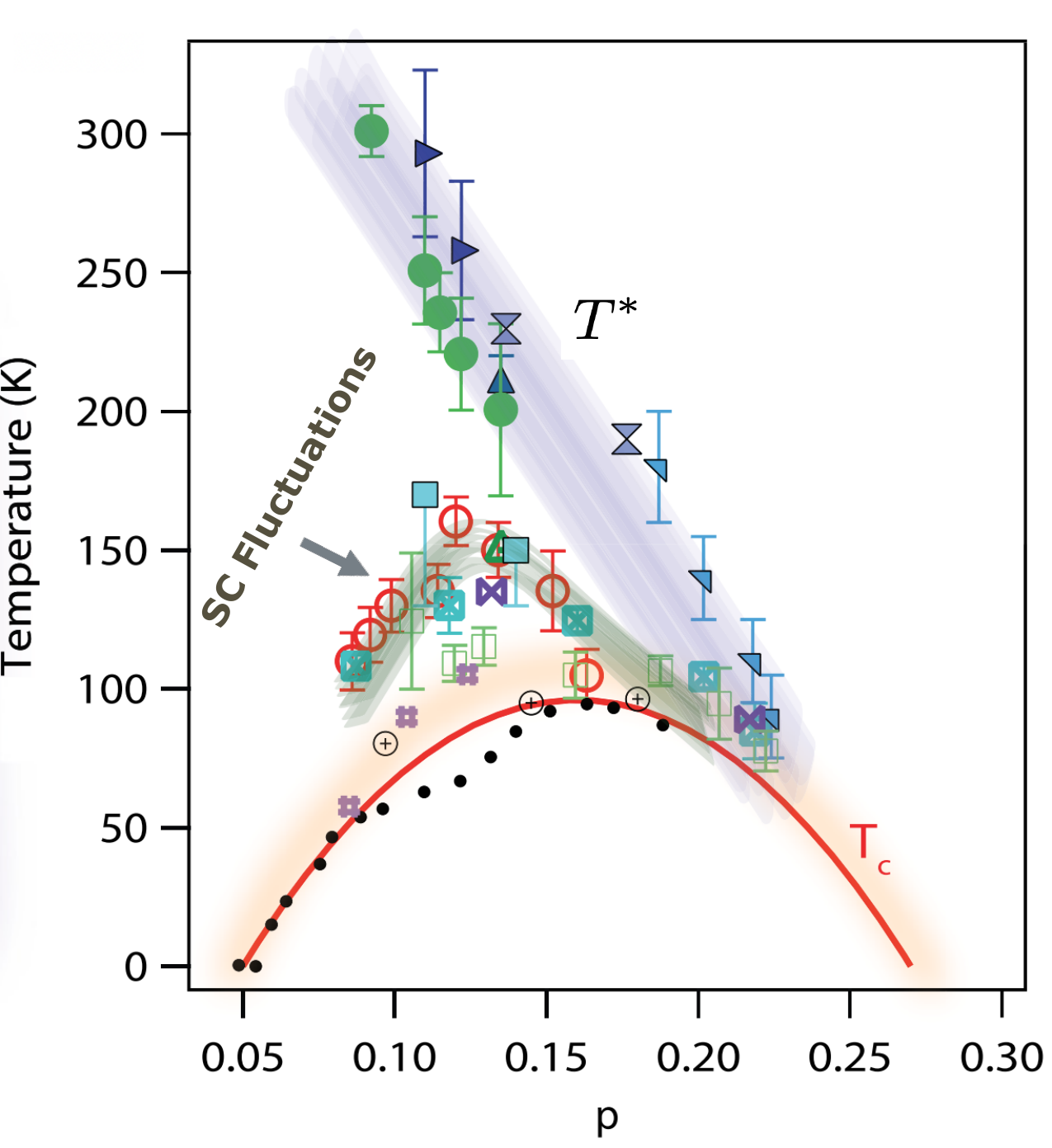}
\caption{Summary of cuprate data~\cite{Vishik2018} and the 
various energy scales which are prototypical in BCS-BEC theory.
The temperature $T^*$ represents the
opening of a fermionic gap and is not directly relevant to bosonic
transport, as this requires a very small pair chemical potential.
Enhanced fluctuation transport of interest in this paper, thus, takes place closer to $T_c$ as indicated
in the figure. This temperature scale is consistent with
the regime in Figure 2b where the bosonic chemical potential 
$|\mupair|$
is sufficiently
small: $|\mupair| \ll  T_c$.}
\label{fig:Fig6}
\end{figure}

\begin{figure}
\includegraphics[width=3.3in,clip]
{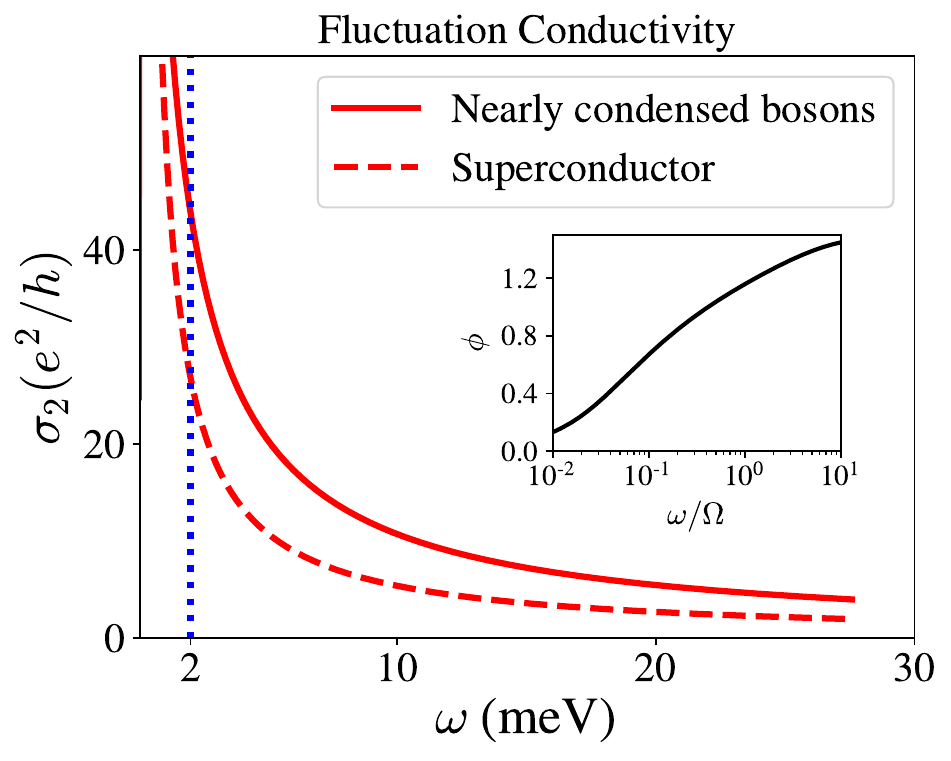}
\caption{This figure shows the behavior of preformed pairs which
contribute to $\sigma_2$, as \textit{equilibrium} superconducting fluctuations
and the contrast with a true ordered superconductor.  Here $T = 1.2 T_c$ and
we presume $T^*/T_c = 15$.
The inset plots the angle $\phi = \tan^{-1} (\frac{\sigma_2}{\sigma_1}$)
and makes a
connection to fluctuation experiments in Refs.~\onlinecite{Corson1999} and \onlinecite{Bilbro2011}. Here $\Omega$ is
arbitrary and chosen to be 1 meV. The blue vertical line shows the lower limit accessible
experimentally.}
\label{fig:Fig5}
\end{figure}

\section{Discussion}

A central goal of this paper was to present a more universal scenario for
the phenomenon of light enhanced ``superconductivity".
The materials where this is observed seem to nicely overlap with those which
are associated with stronger than BCS pairing
\cite{Chen2024}) such as
the particular fulleride-- K$_3$C$_{60}$, and
organic superconductor--$\kappa-\text{BEDT-TTF}$
as well as members of the cuprate and Fe-Se families.

In this context, it should be noted that the ``strong pairing glue",
preformed
pair scenario which we present here can be put in the context of
the widely discussed ``phase fluctuation picture"~\cite{Emery1995}.
In this latter approach it is claimed that ``phonon excitation may
transfer phase coherence to the preformed pairs."
Here we present a more microscopic linkage
which emphasizes the role of fermions: specific phonons help to target those fermions
which participate in the pairing. This, in turn, causes their redistribution to
higher energies and thereby (as in the Eliashberg mechanism~\cite{Eliashberg1970,Tredwell1975,Tredwell1976})
enhances pairing.

While we have emphasized the implications for transport from the bosonic perspective,
this reduction in fermion number will also affect the measured transport directly. In particular,
for the diamagnetic signal, well known paramagnetic contributions
from
Pauli and from the larger~\cite{Li2010} orbital Van Vleck mechanism~\cite{Kubo1956}
will be mostly removed,
thereby leading to even more enhanced diamagnetism than was estimated here.
In this sense, a light induced state should be viewed as a new phase of matter. It is associated with a
removal of a large fraction of the fermions making it different from, say,
the preformed pair state in an equilibrated, underdoped (large $T^*$)
cuprate.

\section{Methods}

\begin{figure}[h]
\includegraphics[width=3.3in,clip]
{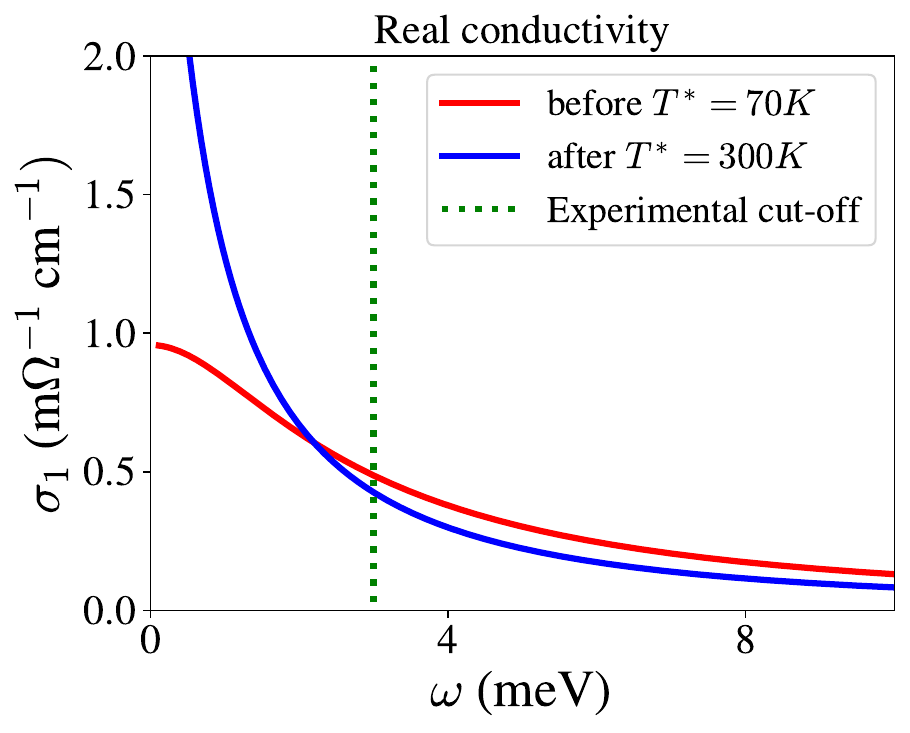}
\caption{This figure plots the real part of the
conductivity before and after exposure to light. Over the range
of measured frequencies one can see that there is very little
change, even less than observed for the imaginary contribution
which is plotted in Fig. \ref{fig:3}. One should note that these
two figures are plotted over different ranges of frequency, however.
}
\label{fig:Fig7}
\end{figure}

\subsection{The ac Conductivity of Preformed Pairs}

Here we focus on pairing fluctuation contributions to the transport,
~\cite{Larkin2005,Tan2004} where the ac conductivity $\sigma(\omega)$
\textit{in the normal state} is given by
\begin{multline}\label{eq:ACconductivity}
{\sigma}_{ij}(\omega)=4\frac{ie^2}{\omega}\int\frac{\dif \vecq}{(2\pi)^d}
\frac{\dif\omega'}{2\pi}v_i(\vect{q})v_j(\vect{q})
\AT(\vect{q}, \omega')b(\omega')\\
\times
Z
\big[\Tmat(\vect{q},\omega^\prime +\omega)
+\Tmat^*(\vect{q},\omega^\prime -\omega)
-\Tmat(\vect{q}, \omega')-\Tmat^*(\vect{q}, \omega^\prime)\big].
\end{multline}
Here
$Z t(\vecq,\omega)$ is the (retarded) pair propagator,
$d$ represents the spatial dimension of the system which we take to be quasi-2d.
$\AT(\vect{q}, \omega) = \text{Re} [2i  Z \Tmat(\vect{q}, \omega)]$
is the boson spectral function and
$v_i(\vect{q}) = q_i/M_\text{B}$ is the boson velocity along the $i$th-direction;
for the quasi-2d system we consider, we take $\sigma(\omega)$ to be the diagonal (and isotropic) in-plane component of $\sigma_{ij}$.

It is important to stress that dominating this transport is the
generally small parameter $|\mupair|$, which sets the scale for
divergences and other features~\cite{Larkin2005}.
Knowing that $\sigma(\omega)$ is governed by a very low temperature scale,
it follows that the imaginary part of the conductivity at moderate frequencies, $\omega \gg |\mupair(T)|$, will
behave quite universally as
$ \sigma_2 (\omega) \approx \frac{ne^2}{m \omega}. $
Here $n/m$ is the effective fermionic number density to mass ratio.

In order to convert to realistic units we assume that under radiation
exposure
the bosonic contribution dominates the ac conductivity and thus will
exhaust the $f$-sum rule. This provides a normalization for
$\sigma_1 (\omega)$ in terms of the measured plasma frequency,
$\omega_p$. Throughout we will use the experimental value for this
plasma frequency, $ \omega_p = 160$ meV, taken from
Refs.~\onlinecite{Dodge2023} and \onlinecite{Dodge2023a}. This is
chosen to be representative of a prototypical material which manifests
light enhanced ``superconductivity" \footnote{Here, in addition it is
  useful to convert conductivity to conventional units via
  $ e^2/(\hbar d) = 1.58 \times 10^5 \Omega^{-1} m^{-1}$ where we
  take~\cite{Corson1999} for a quasi-2d system, with an inter-layer
  spacing $d = 1.54 nm$.}.

Since we are interested in a wide range of frequencies,
the fully self consistent integral expression in Eq.~(\ref{eq:tmatrix}) is cumbersome to use for the
$t$-matrix
appearing in the conductivity equation Eq.~(\ref{eq:ACconductivity}).
%
Thus, for the purposes of transport, we will consider a simple, generic boson propagator
given by
\begin{equation}\label{eq:vertex}
Z\Tmat(\vect{q},\omega) \equiv \left[\omega-\GT(\omega)- \frac{ \vect{q}^2}{2 M_B}
- \mupair  +\frac{i}{2}\FT(\omega)\right]^{-1}.
\end{equation}
The specific details of these self energy parameters are not important and do not affect the conclusions
\footnote{
Here
as in Ref.~\onlinecite{Tan2004},
we take $\GT(\omega), \FT(\omega)  \propto \omega$
with prefactors $ < 1 $ and a smaller prefactor for this second parameter}.

As an additional approximation, it is useful for our purposes to obtain a more analytical expression
for the centrally important pair chemical potential $\mupair$
in terms of $T_c$ and $T^*$. We find our more detailed numerical
calculations in an equilibrium system \cite{Boyack2018}
are reasonably well fitted, as plotted in Fig.~\ref{fig:2} (right panel), by
$-\mupair = \frac{8}{\pi \eta}(T-T_c) + \alpha \frac{(T-T_c)^3}{(T^*-T)^2}$,
with fitted parameters $\eta$ and $\alpha$. Here we take
$\eta \approx 10-20$ and
$\alpha \approx 15$.
Necessarily, $\mupair$ vanishes at $T_c$ and reaches $- \infty$ at $T^*$
where the bosons disappear.

Fig.~\ref{fig:Fig6}
is included to show the various energy scales which appear in a prototypical
BCS-BEC phase diagram~\cite{Chen2024}, here, for concreteness,
plotted for the cuprates.
Panel (a) is the phase diagram shown in Figure 2a which emphasizes
the temperature scale $T^*$. Unlike $T_c$, $T^*$ is monotonic and uniquely
associated with a
distinct superconductor. Importantly, this temperature is not
directly relevant to our transport considerations, as
enhanced conductivity behavior is only to be
expected when the pair chemical potential (shown in Figure 2b) is small, which
is the case much closer to $T_c$.
This important point is made clear by the middle curve of panel (b) in
Fig.~\ref{fig:Fig6}. 
This middle curve underlines the observation, consistent with the present
calculations, that fluctuation transport effects are only apparent
at temperatures well away from $T^*$, and nearer to $T_c$.

\subsection{Comparison with equilibrated superconducting fluctuations}

An important calibration of the conductivity theory here is to compare with
earlier equilibrium data (here for the cuprates) which is presented in two different
references~\cite{Bilbro2011,Corson1999}
addressing the near-$T_c$ fluctuation
contribution to $\sigma(\omega)$ for microwave~\cite{Corson1999}
and THz frequencies~\cite{Bilbro2011}.
A key summary plot focuses on
the angle $\phi = \tan^{-1} (\frac{\sigma_2}{\sigma_1}$)
as a function of frequency, which can be more directly compared.
This is particularly useful as the broad quasi-particle background
was deliberately removed in experiment~\cite{Corson1999} so that
this provides access to the transport properties with negligible 
fermion contributions, much like in the Eliashberg scenario.

Plotted in Fig.~\ref{fig:Fig5} is the fluctuation contribution associated with the pairs,
which can be more directly compared to another 
class of experiments~\cite{Corson1999,Bilbro2011}.
Here we indicate the frequency dependent behavior
of $\sigma_2$ at $T= 1.2T_c$ as
compared with an ideal superconductor (with superfluid fraction $\frac{n_s}{n} = 0.6$).
Indeed, the plot shows that this normal state transport is rather similar to that of
the actual superconductor for temperatures near $T_c$.

The inset plots the angle 
$\phi = \tan^{-1} (\frac{\sigma_2}{\sigma_1})$
and this
behavior is seen to be semi-quantitatively similar to that found experimentally
(see Figure 3 in Ref.\onlinecite{Corson1999} and Figure 3 in Ref.~\onlinecite{Bilbro2011}).
Since we have argued that the fermion contributions to the ac conductivity are negligible in the light-induced preformed pair phase,
these experimental figures serve to nicely characterize this rather new phase of quantum matter.

Finally, for completeness it is useful to present the plot in Fig.~\ref{fig:Fig7} of the Real contribution to the 
conductivity before and after exposure to light.
Note that there is very
little change and very little suppression of the real part of the conductivity over the
range of measured frequencies.
This can be compared with
the behavior observed for the imaginary contribution
which is plotted (over a different frequency scale)
in Fig. \ref{fig:3}, where the change is more significant.

\section{Data availability}
The data analyzed in the current study are available from the author Ke Wang on reasonable request.

\section{Code availability}
The codes used for the current study are available from the author Ke Wang on reasonable request.

\section{Acknowledgment } 

We thank Steve Dodge, Andrea Cavalleri, and Andrew Higginbotham for very helpful discussions
and communications.
Z. W. and Q. C. are supported by the Innovation Program for Quantum Science and Technology (Grant No. 2021ZD0301904).
We also acknowledge the University of Chicago's Research Computing Center for their support of this work.

\section{Author Contributions}
K.L. conceived and supervised the project. K.W. performed the computations. K.W., Q.C. and Z.W. contributed to the acquisition of the data and preparation of figures. All authors have contributed to the interpretation of the data and the drafting as well as the revision of the manuscript.

\section{Competing Interests}
The authors declare no competing interests.

\section{ ADDITIONAL INFORMATION}
Correspondence and requests for materials should be addressed to the authors K. Wang and K. Levin.

\bibliography{Reference1}
\end{document}